\documentstyle[preprint,epsfig,aps]{revtex}

\begin{document}

\draft
%\twocolumn[\hsize\textwidth\columnwidth\hsize\csname@twocolumnfalse%
%\endcsname
%\baselineskip 18pt
%\draft

\title{$\pi^0$ Photo- and Electroproduction at Threshold within a
Dynamical Model}
\author{ S. S. Kamalov\cite{Sabit}, Guan-Yeu Chen and Shin  Nan Yang}
\address{Department of Physics, National Taiwan University, Taipei, Taiwan
10764, Republic of China}
\author{D. Drechsel and L. Tiator}
\address{Institut f\"ur Kernphysik, Universit\"at Mainz, 55099 Mainz,
Germany}

\date{\today}
\maketitle

\begin{abstract}
We show that, within a meson-exchange dynamical model describing
most of the existing pion electromagnetic production data up to
the second resonance region, one is also able to obtain a good
agreement with the $\pi^0$ photo- and electroproduction data near
threshold. The potentials used in the model are derived from an
effective chiral Lagrangian. The only sizable discrepancy between
our results and the data is in the $P-$wave amplitude
$P_3=2M_{1+}+M_{1-}$ where our prediction underestimate the data
by about 20\%. In the case of $\pi^0$ production, the effects of
final state interaction in the threshold region are nearly
saturated by single charge exchange rescattering. This indicates
that in ChPT it might be sufficient to carry out the calculation
just up to one-loop diagrams for threshold neutral
pion production.\\

\end{abstract}
%]
%\pacs{PACS numbers: 13.60.Le, 14.20.GK \\
% KEYWORDS: pion, photoproduction, electroproduction, Low Energy Theorem,
% threshold }

%\section{INTRODUCTION}

Chiral perturbation theory (ChPT)
%is now widely accepted as the "basic" theory
provides us with a systematic scheme to describe the low energy
interactions of Goldstone bosons among themselves and with other
hadrons, because it is based on a low energy effective field
theory  respecting the symmetries of QCD, in particular chiral
symmetry. There is generally good agreement between the ChPT
predictions and experiments \cite{chiral96}. One case which has
been very intensively studied is $\pi^0$ electromagnetic
production of neutral pions near threshold where very precise
measurements
\cite{Beck90,Fuchs96,Bergstrom,NIKHEF,Distler,Bernstein97,Schmidt}
have been performed and the ChPT calculation to one loop $O(p^3)$
($O(p^4)$ in the case of photoproduction) has been carried out in
the heavy baryon formulation \cite{Bernard91,Bernard}. Nice
agreement between theory and experiment was reached not only for
the $S-$wave multipoles $E_{0+}$ and $ L_{0+}$ but for the
$P-$wave amplitudes \cite{Schmidt,Bernard} as well.

As in ChPT, meson-exchange models also start from an effective
chiral Lagrangian. However, they differ from ChPT in the approach
to calculate the scattering amplitudes. In ChPT, crossing symmetry
is maintained in the perturbative field-theoretic calculation, and
the agreement with low energy theorems and the data is to be
expected as long as the series is well convergent. In
meson-exchange models, the effective Lagrangian is used to
construct a potential for use in the scattering equation. The
solutions of the scattering equation will include rescattering
effects to all orders and thereby unitarity is ensured, while
crossing symmetry is violated. Such models
\cite{pearce,lee91,hung,gross93,SL96,afnan99,tjon00} have been
able to provide a good description of $\pi N$ scattering lengths
and phase shifts in $S, P,$ and $D$ waves up to 600 MeV pion
laboratory kinetic energy.

Meson-exchange models have been constructed for pion
electromagnetic production as well
\cite{SL96,NBL,yang91,hsiao,gross96} and good agreement with the
data has also been achieved up to 1300 MeV total $\pi N$ c.m.
energy. However, the predictive power of the meson-exchange model
for electromagnetic pion production near threshold has not been
fully explored even though the importance of final state
interaction (FSI) for threshold $\pi^0$ photoproduction had been
demonstrated in several dynamical model studies
\cite{lee91,yang89,nbl90,pearce91} prior to one-loop calculations
of ChPT \cite{Bernard91}.

In this paper we will use the dynamical model (DM) recently
developed in Ref. \cite{KY99} where the dominant FSI effects in
the nonresonant contributions at threshold  are evaluated using
the $\pi N$ meson-exchange model developed in Ref. \cite{hung}.
Contributions which are related to the excitation of resonances
are considered phenomenologically using standard Breit-Wigner
forms. In our previous work \cite{KY99} such an approach gives an
excellent description of pion photo- and electroproduction in the
first resonance region. Here we apply this model to these
reactions in the threshold region and compare our predictions with
the recent experimental data
\cite{Fuchs96,Bergstrom,NIKHEF,Distler,Bernstein97,Schmidt} for
the $S-$ and $P-$wave multipoles and cross sections, and with the
result of ChPT.

%\section{Formalism}

In the dynamical model for  pion photo- and electroproduction
\cite{Tanabe,Yang85}, the t-matrix is given as
\begin{eqnarray}
t_{\gamma\pi}(E)=v_{\gamma\pi}+v_{\gamma\pi}g_0(E)\,t_{\pi
N}(E)\,,\label{eq:tmatrix}
\end{eqnarray}
where $v_{\gamma\pi}$ is the $\gamma\pi$ transition potential,
$g_0$ and $t_{\pi N}$ are the $\pi N$ free propagator and
$t-$matrix, respectively, and $E$ is the total energy in the c.m.
frame. In the present study, the matrix elements $t_{\pi N}$ are
obtained in a meson-exchange $\pi N$ model \cite{hung}
constructed in the Bethe-Salpeter formalism and solved within
Cooper-Jennings reduction scheme \cite{CJ89}. Both $v_{\pi N}$ and
$v_{\gamma\pi}$  are derived from an  effective Lagrangian
containing Born terms as well as $\rho-$ and $\omega-$exchange in
the $t-$channel \cite{Olsson,MAID98}. For pion electroproduction
we restore gauge invariance by the substitution,
\begin{eqnarray}
 J_{\mu} \rightarrow J_{\mu}  - k_{\mu}\frac{k\cdot J}{k^2}\,,
\end{eqnarray}
where $J_{\mu}$ is the electromagnetic current corresponding to
the background  contribution of $v_{\gamma\pi}$.

For the physical multipoles in channel $\alpha=\{l,j\}$, Eq.
(\ref{eq:tmatrix}) gives~\cite{Yang85}
\begin{eqnarray}
t_{\alpha}(q_E,k)=\exp{(i\delta_{\alpha})}\,\cos{\delta_{\alpha}}
\left[ v_{\alpha}(q_E,k) + P\int_0^{~} dq'
\frac{R_{\alpha}(q_E,q')\,v_{\alpha}(q',k)}{E(q_E)-E(q')}\right]\,,
\label{eq:Tback}
\end{eqnarray}
where $\delta_{\alpha}$ and $R_{\alpha}$ are the $\pi N$ phase
shift and  reaction matrix, in channel $\alpha$, respectively,
$q_E$ is the pion on-shell momentum and $k=\mid {\bf k}\mid$ the
photon momentum. In order to ensure the convergence of the
principal value integral, we introduce a dipole-like off-shell
form factor characterizing the finite range aspect of the
potential,
$f(q,q_E)=\left(\Lambda^2+q_E^2)^2/(\Lambda^2+q^2\right)^2.$
%\begin{eqnarray}
%   f(q,q_E)=\left(\frac{\Lambda^2+q_E^2}{\Lambda^2+q^2}\right)^2.
%\end{eqnarray}
The value for the cut-off parameter, $\Lambda=440$ MeV, was
obtained in our previous work from an analysis of the
$\Delta(1232)$ resonance multipole $M_{1+}^{(3/2)}$ over a wide
energy range.

%\subsection{$\pi^0$ photoproduction}

For $\pi^0$ photoproduction, we first calculate the multipole
$E_{0+}$ near threshold by solving the coupled channels equation
within a basis with physical pion and nucleon masses. The coupled
channels equation leads to the following expression for the pion
photoproduction t-matrix in the $\pi^0p$ channel:
\begin{eqnarray}
t_{\gamma\pi^0}(E)& = & v_{\gamma\pi^0}(E)+v_{\gamma\pi^0}(E)\,
g_{\pi^0 p}(E)\,t_{\pi^0 p\rightarrow \pi^0 p}(E) \nonumber\\& + &
v_{\gamma\pi^+}(E)\, g_{\pi^+ n}(E)\,t_{\pi^+ n\rightarrow \pi^0
p}(E)\,, \label{eq:coupled}
\end{eqnarray}
where $t_{\pi^0 p\rightarrow \pi^0 p}$ and $t_{\pi^+ n\rightarrow
\pi^0 p}$ are the $\pi N$  t-matrices in the elastic and charge
exchange channels, respectively. They are obtained by solving the
coupled channels equation for $\pi N$ scattering using the
meson-exchange model~\cite{hung}. Results for $Re\,E_{0+}$
obtained from Eq. (\ref{eq:coupled}) are given in Fig. 1, where
the FSI contributions from  the elastic and charge exchange
channels (second and third term in Eq. (\ref{eq:coupled})), are
shown by the short-dashed and dash-dotted curves, respectively,
while the dotted curve corresponds to the LET results, without the
inclusion of FSI. Our results clearly indicate that practically
all of the final state interaction effects originate from the
$\pi^+ n$ channel. Note that the main contribution stems from the
principal value integral of Eq. (\ref{eq:coupled}).

In the coupled channels approach considered above,  the  $t_{\pi
N}$ matrix  contains the effect of $\pi N$ rescattering to all
orders. However, we have indeed found that only the first order
rescattering contribution, i.e. the one-loop diagram, is
important. This result is obtained by replacing the $\pi N$
scattering t-matrix in Eq. (\ref{eq:coupled}) with the
corresponding potential $v_{\pi N}$. As can be seen in Fig. 1, the
thus obtained results given by the long-dashed curve, differ from
the full calculation (solid curve) by 5\% only. This indicates
that the one-loop calculation in ChPT could be a reliable
approximation for $\pi^0$ production in the threshold region.

If the  FSI effects are evaluated with the assumption of isospin
symmetry (IS), i.e., with averaged masses in the free pion-nucleon
propagator, the energy dependence in $Re\, E_{0+}$ in the
threshold region is very smooth. Below $\pi^+$ threshold the
strong energy dependence (cusp effect)
\cite{Bernstein97,Bernstein98} only appears because of the pion
mass difference and, as we have seen above, is related to the
coupling with the  $\pi^+n$ channel. In most calculations, the
effects from the pion mass difference below the $\pi^+$ production
channel are taken into account by using the K-matrix approach
\cite{Bernard,Laget},
\begin{eqnarray}
Re\, E_{0+}^{\gamma\pi^0} = Re\, E_{0+}^{\gamma\pi^0}(IS)
 -  a_{\pi N}\,\omega_c\,Re\,
E_{0+}^{\gamma\pi^+}(IS)\sqrt{1-\frac{\omega^2}{\omega^2_c}}\,,
\label{eq:kmatr}
\end{eqnarray}
where $\omega$ and $\omega_c$ are the $\pi^0$ and $\pi^+$ c.m.
energies corresponding to  $E=E_p + E_{\gamma}$ and
$m_n+m_{\pi^+}$, respectively, and $a_{\pi N}=0.124/m_{\pi^+}$ is
the pion charge exchange threshold amplitude. $E_{0+}^{\gamma
\pi^{0,+}}(IS)$ is the $\pi^{0,+}$ photoproduction amplitude
obtained with the assumption of isospin symmetry (IS), i.e.,
without the pion mass difference in Eq. (\ref{eq:Tback}). Such an
approximation is often used in the data analysis in order to
parametrize the $E_{0+}$ multipole below $\pi^+n$ threshold in the
form of $E_{0+}(E)=a + b \sqrt{1-(\omega/\omega_c)^2}$. In Fig. 2
the results obtained from such an approximation scheme are
represented by the solid curve and compared to the exact coupled
channels (dashed curve) and ChPT calculations (dash-dotted curve)
\cite{Bernard91}. We see that above  $\pi^+$ threshold the
difference between the results obtained with physical and isospin
bases is very small. This is consistent with the finding of
Ref.\cite{Laget}. At lower energies the difference becomes visible
only very close to $\pi^0 p$ threshold where the two approaches
differ by about 10\%. In general we can conclude that the simple
expression of Eq. (\ref{eq:kmatr}) is a good approximation for the
pion-mass difference effect, and in the following calculations we
will use this option to analyze the experimental data. Note that
the correct threshold dependence of the imaginary part can be
obtained from the Fermi-Watson theorem if in the threshold region
the $\pi N$ phase shift is taken as a linear function of the
$\pi^+$ momentum, i.e., vanishing as $q_{\pi^+}\rightarrow 0$.

In  Fig. 3, we compare the predictions of our model for the
differential cross section with recent photoproduction data from
Mainz \cite{Fuchs96,Schmidt}. The dotted and solid curves are
obtained without and with FSI effects, respectively. It is seen
that both off-shell pion rescattering and cusp effects
substantially improve the agreement with the data. This indicates
that our model gives reliable predictions also for the threshold
behaviour of the $P$-waves without any additional arbitrary
parameters. As an example, numerical values for $E_{0^+}$ (in
units of $10^{-3}/m^3_{\pi^+}$) and $P$-wave multipoles (in units
of $10^{-3}q/m^3_{\pi^+}$) at $\pi^0$ threshold are given in Table
I. For the $P$-wave multipoles we give values for the following
linear combinations:
$P_1=3E_{1+}+M_{1+}-M_{1-},\,P_2=3E_{1+}-M_{1+}+M_{1-}$ and
$P_3=2M_{1+}+M_{1-}$.

The contributions of  Born terms, vector meson exchange, FSI and
resonances ($S_{11}(1535),\,\Delta(1232)$ and $P_{11}(1440)$) are
listed in Table I to show their relative importance.  One observes
that $\rho-$ and mostly $\omega-$exchange give important
contributions to the $P$-wave amplitudes, especially to $P_3$,
where Born terms contribute only 2\%. Half of $P_3$ comes from the
vector meson exchanges and the rest from FSI and resonance
contributions. Due to the large contribution from these three
mechanisms, $P_3$ becomes comparable to $P_1$ and $P_2$. Very
close to threshold, our model predicts $\mid P_3/P_2 \mid \simeq
0.9$, and a small but negative value for the photon asymmetry
$\Sigma(\theta_\pi = 90^\circ)\sim \mid P_3 \mid^2 - \mid P_2
\mid^2$ at fixed pion c.m. angle $\theta_\pi = 90^\circ$. As shown
by the solid curve in Fig. 4, our prediction for
$\Sigma(\theta_\pi = 90^\circ)$ first tends to more negative
values before bending over and becoming positive at large photon
energies. It was found in Ref. \cite{HDT} that in the threshold
region this observable is very sensitive to the $M_{1-}$ multipole
which strongly depends on the details of the low energy behavior
of Roper resonance, vector meson and  FSI contributions.
Therefore, a slight modification of one or all of these mechanisms
can drastically change the photon asymmetry. As an illustration,
our prediction for the energy dependence and angular distribution
of $\Sigma(\theta_\pi)$ obtained with a 15\% reduction of the
$M_{1-}$ multipole, is shown by dashed curves in Fig. 4.  This
small modification of the low energy tail of the Roper resonance
leads to positive photon asymmetries at all energies!

In Table I, the ChPT predictions and  the experimental values
extracted from recent TAPS polarization measurements
\cite{Schmidt} are listed for comparison. Our predictions are in
good overall agreement with the ChPT predictions \cite{Bernard}
and the TAPS results. However, there is a $15\% - 20\%$ difference
in $P_3$ which leads to an underestimation of our result for the
photon asymmetry, as shown in Fig. 4. Note that, in contrast to
our model, $P_3$ is essentially determined by a low energy
constant in ChPT.

%\subsection{$\pi^0$ electro-photoproduction}

Pion electroproduction provides us with information on the $Q^2 =
-k^2$ dependence of the transverse $E_{0+}$ and longitudinal
$L_{0+}$ multipoles in the threshold region. The "cusp" effects in
the $L_{0+}$ multipole is taken into account in a similar way as
in the case of $E_{0+}$,
\begin{eqnarray}
Re\, L_{0+}^{\gamma\pi^0} = Re\, L_{0+}^{\gamma\pi^0}(IS)
- a_{\pi N} \,\omega_c\,Re\,
L_{0+}^{\gamma\pi^+}(IS)\,\sqrt{1-\frac{\omega^2}{\omega^2_c}}\,,\label{eq:L-Kmatrix}
\end{eqnarray}
where all the multipoles are functions of total c.m. energy $E$
and virtual photon four-momentum squared $Q^2$. It is known that
at threshold, the $Q^2$ dependence is given mainly by the Born
plus vector meson contributions in $v_{\gamma\pi}^B$, as described
in Ref.~\cite{MAID98}.

The validity of the approximation made in Eq. (\ref{eq:L-Kmatrix})
is confirmed in Fig. 5 where the results of K-matrix approximation
(solid lines) and full calculation (dash-dotted lines) agree with
each other within a few percent. In Fig. 5 we also show our
results for the cusp and FSI effects in the $E_{0+}$ and $L_{0+}$
multipoles for $\pi^0$ electroproduction at $Q^2=0.1$ (GeV/c)$^2$,
along with the results of the multipole analysis from
NIKHEF\cite{NIKHEF} and Mainz\cite{Distler}. Note that results of
both groups were obtained using the $P$-wave predictions given by
ChPT. However, there exist substantial differences between the
$P-$wave predictions of ChPT and our model at finite $Q^2$, as
presented in Table II. To understand the consequence of these
differences,  we have made a new analysis of the Mainz
data\cite{Distler} for the differential cross sections, using our
DM prediction for the $P$-wave multipoles instead. The $S$-wave
multipoles extracted this way  are also shown in Fig. 5 by solid
circles and listed in Table III. We see that the results of such a
new analysis gives a $E_{0+}$ multipole closer to the NIKHEF data
and in better agreement with our dynamical model prediction.
However, the results of our new analysis for the longitudinal
$L_{0+}$ multipole stay practically unchanged from the values
found in the previous analyses. Note that the dynamical model
prediction for $L_{0+}$  again agrees much better with the NIKHEF
data.

In Figs. 6 and 7, our model predictions (dashed curves) are
compared with the Mainz experimental data\cite{Distler} for the
unpolarized cross sections $d\sigma/d\Omega=d\sigma_T/d\Omega +
\epsilon\,d\sigma_L/d\Omega$, and for the longitudinal-transverse
cross section $d\sigma_{TL}/d\Omega$. Overall, the agreement is
good. The solid curves are the results of our best fit at fixed
energies (local fit) obtained by varying only the $E_{0+}$ and
$L_{0+}$ multipoles. We have found that the differences between
the solid and dashed curves  in Figs. 6 and 7 are mostly due to
the difference in the $L_{0+}$ multipole (see also Fig. 5).

Finally we discuss a new version (hereafter called MAID)
\cite{yang01} of the unitary isobar model developed at Mainz
(hereafter called MAID98) \cite{MAID98}, which  is currently being
intensively used for the analysis of pion photo- and
electroproduction data. In this model the FSI effects are taken
into account using the K-matrix approximation, namely without the
inclusion of off-shell pion rescattering contributions (principal
value integral) in Eq. (\ref{eq:Tback}). As a result, the $S$-,
$P$-, $D$- and $F$-waves of the background contributions are
defined in MAID as
\begin{eqnarray}
t_{\alpha}^B(MAID)=\exp{(i\delta_{\alpha})}\,\cos{\delta_{\alpha}}
\,v_{\alpha}^B(q_E,k).
\label{eq:Tmaid}
\end{eqnarray}

However, as we have found above, dynamical model calculations show
that pion off-shell rescattering is very important at low pion
energies. The prediction of MAID for $\pi^0$ photoproduction at
threshold, represented by the dotted curves in Figs. 1 and 2, lies
substantially below the data. It turns out that it is possible to
improve MAID, in the case of $\pi^0$ production at low energies,
by introducing a phenomenological term and including the cusp
effect of Eq. (\ref{eq:kmatr}). In this extended version of MAID,
we write the $E_{0+}(\pi^0p)$ multipole as
\begin{eqnarray}
Re\, E_{0+}^{\gamma\pi^0}  =  Re\, E_{0+}^{\gamma\pi^0}(MAID98) +
E_{cusp}(W,Q^2) + & E_{corr}(W,Q^2)\,, \label{eq:Emaid}
\end{eqnarray}
where
\begin{eqnarray}
E_{cusp}= - a_{\pi N} \,\omega_c\,Re\,
E_{0+}^{\gamma\pi^+}(MAID98)\,\sqrt{1-\frac{\omega^2}{\omega^2_c}}\,.
\label{eq:cusp}
\end{eqnarray}
The phenomenological term $E_{corr}$ which emulates the pion
off-shell rescattering corrections (or pion-loop contribution in
ChPT) can be parameterized in the form
\begin{eqnarray}
E_{corr}(W,Q^2)= \frac{A}{(1 + B^2q^2_{\pi})^2}\,F_D(Q^2)\,,
\label{eq:corr}
\end{eqnarray}
where $F_D$ is the standard nucleon dipole form factor. The
parameters $A$ and $B$ are obtained by fitting to the low energy
$\pi^0$ photoproduction data: $A= 2.01 \times 10^{-3}/m_{\pi^+}$
and $B=0.71 \, fm$.

%\section{CONCLUSION}

In summary, we have shown that within a meson-exchange dynamical
model \cite{KY99}, one is able to describe photo- and
electroproduction in the threshold region in good agreement with
the data. The model has been demonstrated \cite{yang01,sabit01} to
give a good description of most of the existing pion
electromagnetic production data up to the second resonance region.
The success of such a model at intermediate energies is perhaps
not surprising since unitarity plays an important role there.
However, it is not {\it a priori} clear that our model should also
work well near threshold, even though we do start from an
effective chiral Lagrangian. In principle, crossing symmetry is
violated and the well-defined power counting scheme in ChPT is
lost by rescatterings. As a matter of fact, previous similar
attempts have failed \cite{yang89,pearce91}. It is easy to
understand the difference between our present calculation and the
results of Ref. \cite{yang89} by the fact that the off-shell
behavior of the $\pi N$ models used \cite{hung} are quite
different from each other. The difference between our results and
those of Ref. \cite{pearce91} probably arises, in large part, from
different off-shell prescriptions used for the transition
potential $v_{\gamma\pi}$, because the meson-exchange $\pi N$
model used in Ref. \cite{pearce91}  is very similar to the one
used in this study.  On the other hand, meson-exchange models
\cite{afnan99,tjon00} have also been shown to give a good
description of low energy $\pi N$ data, in addition to an
excellent agreement with the data at higher energies. It is
therefore assuring that similar success can also be achieved for
the pion electromagnetic production.

The largest discrepancy between our results and the data is in the
$P_3$ amplitude where our prediction underestimates the data by
about 20\%. As a consequence, our prediction for the photon
asymmetry has the opposite sign as observed in the experiment
\cite{Schmidt}. However, we have found that in the threshold
region, the photon asymmetry is very sensitive to many ingredients
of the theory, e.g., vector mesons, FSI and Roper resonance
contributions, a fact that deserves further studies.

Finally, we found that the effects of final state interaction in
the threshold region and in the case of $\pi^0$ production, are
nearly saturated by  the single rescattering term. Therefore, the
existing one-loop calculations in ChPT  could be a good
approximation to threshold neutral pion production.

\acknowledgements

We would like to thank Drs. R. Beck, M. Distler, H. Merkel, and A.
Schmidt for helpful discussions. S.S.K. and L.T. are grateful to
the Physics Department of the NTU for the hospitality extended to
them during their visits. This work is supported in part by the
National Science Council/ROC under grant NSC 89-2112-M002-078, by
the Deutsche Forschungsgemeinschaft (SFB 443) and by a joint
project NSC/DFG TAI-113/10/0.

\begin{table}[htbp]
\begin{center}
\begin{tabular}{|c|cccc|c||c|c|}
       &  Born    &  $\omega+\rho$ & FSI     &  res.  &  tot.   & ChPT & Exp. \\
\hline
$E_{0+}$ & -2.46 &   0.17  &  1.06  &  0.07 & -1.16 & -1.16 & $-1.33\pm 0.11$\\
\hline
$P_{1} $ &  9.12 &  -0.35  &  0.15  &  0.38 &  9.30 &  9.14 & $9.47\pm 0.37$\\
$P_{2} $ & -8.91 &   0.21  & -1.32  & -0.13 & -10.15 & -9.7 & $-9.46\pm 0.39$\\
$P_{3} $ &  0.18 &   4.61  &  3.36  &  1.20 &  9.35 &  10.36 & $11.48\pm 0.41$\\
\end{tabular}
\end{center}
\caption{ $E_{0+}$ (in units $10^{-3}/m_{\pi^+}$) at threshold and
$P_1,\,P_2$ and $P_3$ (in units $10^{-3}q/m^2_{\pi^+}$).
Contributions of Born terms (Born), vector mesons (
$\omega+\rho$), pion rescattering (FSI) and resonances (res.) are
shown separately. The predictions of ChPT and recent experimental
values are taken from Ref.\protect\cite{Bernard} and
Ref.\protect\cite{Schmidt}.}
\end{table}
\begin{table}[htbp]
\begin{center}
\begin{tabular}{|c|cc|cc|cc|}
 $Q^2$ (GeV/c)$^2$ & 0.0 & 0.0 & 0.05  & 0.05 & 0.1 & 0.1\\
\hline
Model    &  DM    &  ChPT    & DM     & ChPT  &  DM   & ChPT  \\
\hline
$E_{0+}$ & -0.92 &  -0.96  &  1.08  &  0.45 &  2.48 &  1.60 \\
$M_{1+}$ &  1.11 &   1.19  &  1.70  &  2.16 &  1.93 &  2.74 \\
$M_{1-}$ & -0.59 &  -0.49  & -0.96  & -0.72 & -1.07 & -0.74 \\
$E_{1+}$ & -0.02 &  -0.02  & -0.03  & -0.01 & -0.03 & -0.01 \\
\hline
$L_{0+}$ & -2.71 &  -1.61  & -1.74  & -1.52 & -1.13 & -1.31 \\
$L_{1-}$ & -0.14 &  -0.52  & -0.05  & -0.60 & -0.01 & -0.51 \\
$L_{1+}$ & -0.01 &  -0.01  & -0.01  & -0.02 & -0.01 & -0.02 \\
\end{tabular}
\end{center}
\caption{ Comparison of the $S$- and $P$-wave multipoles (in units
$10^{-3}/m_{\pi^+}$) for the  $p(\gamma^*,\pi^0)p$ reaction at
$Q^2$=0, 0.05 and 0.10 (GeV/c)$^2$, obtained in our model and ChPT
\protect\cite{Bernard} at $\Delta W$= 2.5 MeV.}
\end{table}
\begin{table}[htbp]
\begin{center}
\begin{tabular}{|c|ccc|}
$\Delta W$ &  $E_{0+}$ & $L_{0+}$ & $\chi^2$/d.o.f. \\
\hline
    0.5    & $ 2.28 \pm 0.36 $ & $ -1.34 \pm 0.06 $ & 1.49 \\
Ref.\protect\cite{Distler}
           & $ 1.96 \pm 0.33 $ & $ -1.42 \pm 0.05 $ & 1.29 \\
\hline
    1.5    & $ 2.43 \pm 0.21 $ & $ -1.38 \pm 0.04 $ & 1.26 \\
Ref.\protect\cite{Distler}
           & $ 1.82 \pm 0.19 $ & $ -1.41 \pm 0.03 $ & 1.19 \\
\hline
    2.5    & $ 2.98 \pm 0.20 $ & $ -1.38 \pm 0.04 $ & 1.60 \\
Ref.\protect\cite{Distler}
           & $ 2.12 \pm 0.17 $ & $ -1.36 \pm 0.05 $ & 1.68 \\
\hline
    3.5    & $ 2.61 \pm 0.22 $ & $ -1.39 \pm 0.03 $ & 1.70 \\
Ref.\protect\cite{Distler}
           & $ 1.52 \pm 0.18 $ & $ -1.27 \pm 0.03 $ & 1.84 \\
\end{tabular}
\end{center}
\caption{ Values of the $E_{0+}$ and $L_{0+}$  (in units
$10^{-3}/m_{\pi^+}$) for the  $p(\gamma^*,\pi^0)p$ reaction  at
$Q^2$=0.10 (GeV/c)$^2$, obtained after the local fit to the
differential cross sections measured in
Ref.\protect\cite{Distler}.}
\end{table}

%%%%%%%%%%%%%%   Fig. 1 ****************
\begin{figure}[h]
\centerline{\epsfig{file=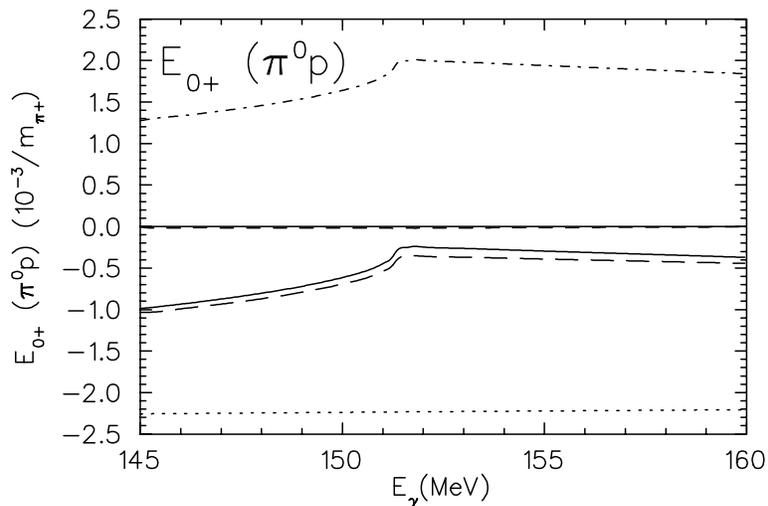, width=10 cm} } \caption{
Real part of the $E_{0+}$ multipole for $\gamma p\rightarrow \pi^0
p$. The dotted curve is the result obtained without FSI, the
dash-dotted and short-dashed curves are the FSI contributions from
charge exchange $\pi^+n$ and elastic $\pi^0 p$ channel,
respectively. The solid and long-dashed curves are the total
results (see  Eq. (\protect\ref{eq:coupled})) obtained with the
full matrix $t_{\pi N}$ and with  $t_{\pi N}$ replaced by $v_{\pi
N}$, respectively.}
\end{figure}

\newpage

%%%%%%%%%%%%%%   Fig. 2 ****************
\begin{figure}[h]
\centerline{\epsfig{file=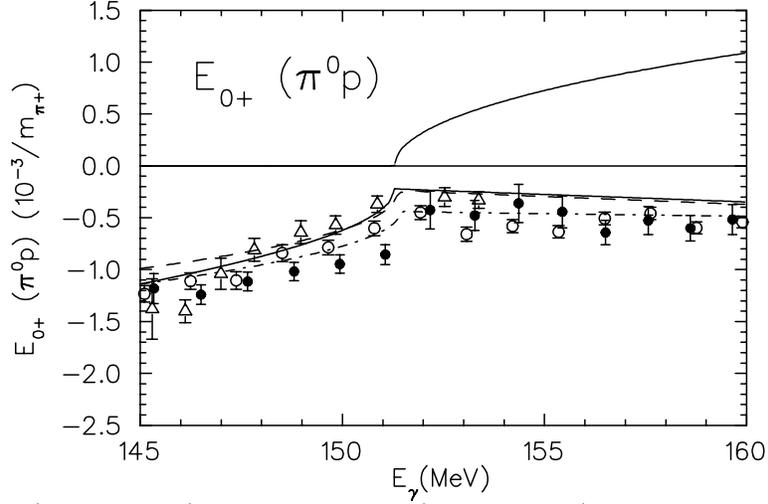, width=10 cm} } \caption{
Real (lower panel) and imaginary (upper panel) parts of the
$E_{0+}$ multipole for $\gamma p\rightarrow \pi^0 p$. The dashed
and solid curves are the full results obtained without and with
isospin symmetry assumption, respectively. In the latter case, the
pion mass difference effect is taken into account using Eq.
(\protect\ref{eq:kmatr}). The dash-dotted curve is the result of
ChPT \protect\cite{Bernard}. Data points from Ref.
\protect\cite{Fuchs96}($\triangle $), Ref.
\protect\cite{Bergstrom}($\bullet$), and Ref.
\protect\cite{Schmidt}($\circ$).}
\end{figure}

%%%%%%%%%%%%%%   Fig. 3 ****************
\begin{figure}[h]
\centerline{\epsfig{file=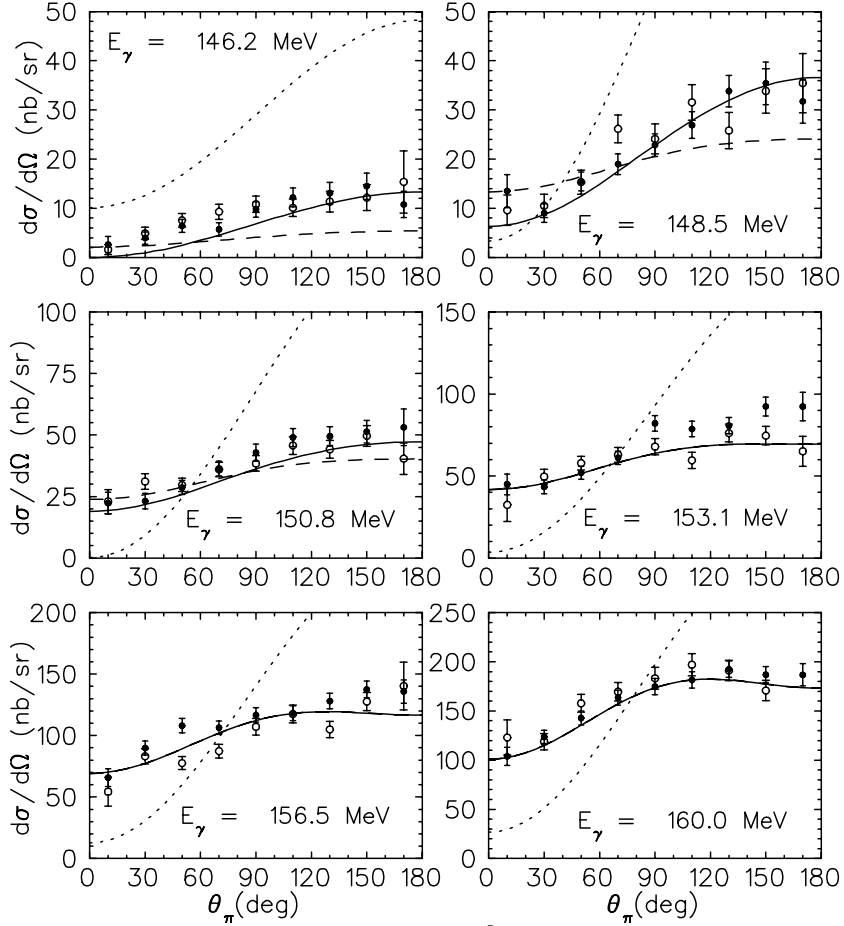, width=11 cm} } \caption{
Differential cross sections  for $\gamma p\rightarrow \pi^0 p$.
 Dotted and dashed curves: the results obtained
without and with FSI contributions using isospin symmetry. Solid
curves: final result including pion mass difference effects.
Experimental data from Ref. \protect\cite{Fuchs96} ($\bullet$) and
Ref.\protect\cite{Schmidt} ($\circ$).}
\end{figure}

%%%%%%%%%%%%%%   Fig. 4 ****************
\begin{figure}[h]
\centerline{\epsfig{file=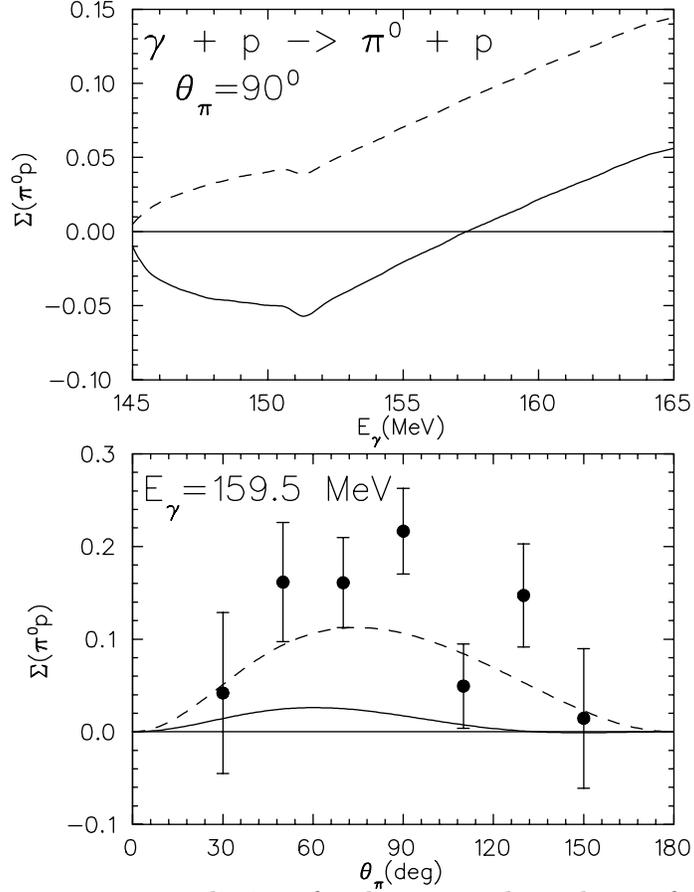, width=9 cm} } \caption{
Solid curves are our predictions for the energy dependence of the
photon asymmetry $\Sigma$ at $\theta_{\pi}=90^0$(upper panel) and
its angular distribution at $E_{\gamma}=159.5$ MeV (lower panel)
in $\gamma p\rightarrow \pi^0 p$. Dashed curves are results
obtained with a 15\% reduction of the $M_{1-}$ multipole in the
model. Experimental data from Ref.\protect\cite{Schmidt}.}
\end{figure}

%%%%%%%%%%%%%%   Fig. 5 ****************
\begin{figure}[h]
\centerline{\epsfig{file=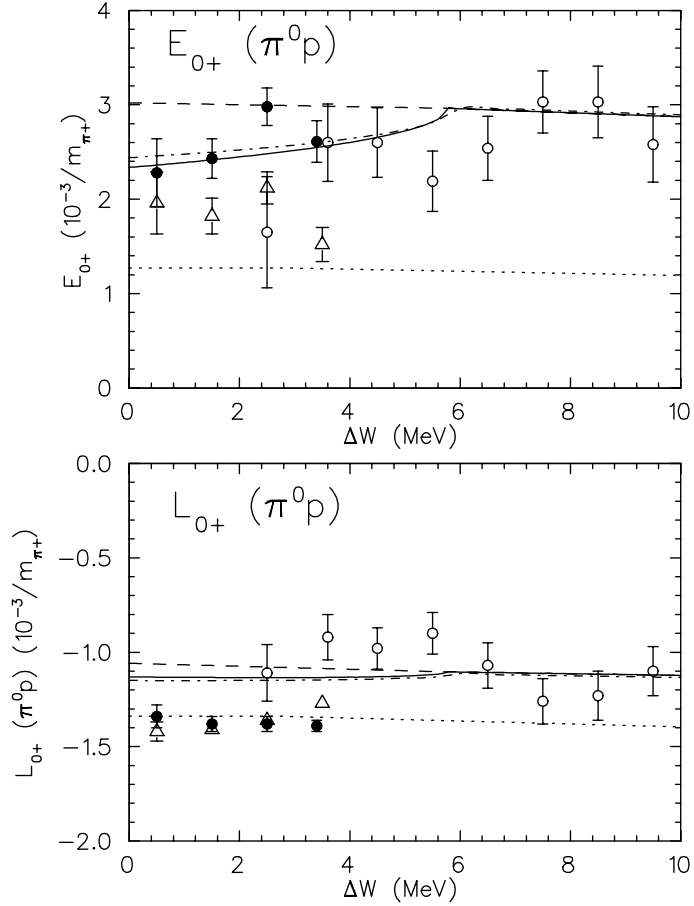, width=9 cm} } \caption{ Real
parts of $E_{0+}$ (upper panel) and  $L_{0+}$ (lower panel) for
$ep \rightarrow e'\pi^0 p$ at $Q^2$=0.1 (GeV/c)$^2$. Dash-dotted
curves are the results of full calculations obtained using Eq. (4)
without the assumption of isospin symmetry. Notations for other
curves are the same as in Fig. 3. Data points from
Ref.\protect\cite{NIKHEF}($\circ$) and
Ref.\protect\cite{Distler}($\triangle$). The results of the
present work obtained by using the $P$-waves of our model are
given by ($\bullet$).}
\end{figure}

%%%%%%%%%%%%%   Fig. 6 ****************
\begin{figure}[h]
\centerline{\epsfig{file=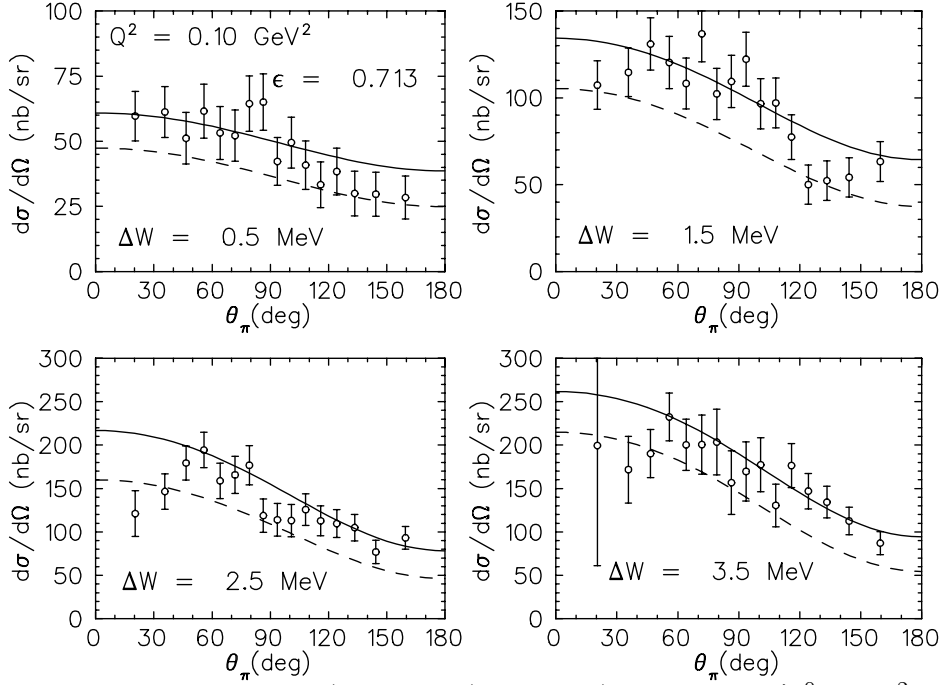, width=9 cm, angle=90 } }
\caption{ Angular distribution $d\sigma/d\Omega=d\sigma_T/d\Omega
+ \epsilon\,\sigma_L/d\Omega$ for $ep \rightarrow e'\pi^0 p$ at
$Q^2$=0.1 (GeV/c)$^2$ and $\epsilon=0.713$, at different total
c.m. energies  $\Delta W=W-W_{thr}^{\pi^0p}$. Dashed curves are
predictions of our model. Solid curves are the results of our
local fit with fixed $p$-waves. Experimental data from
Ref.\protect\cite{Distler}.}
\end{figure}

%%%%%%%%%%%   Fig. 7 ****************
\begin{figure}[h]
\centerline{\epsfig{file=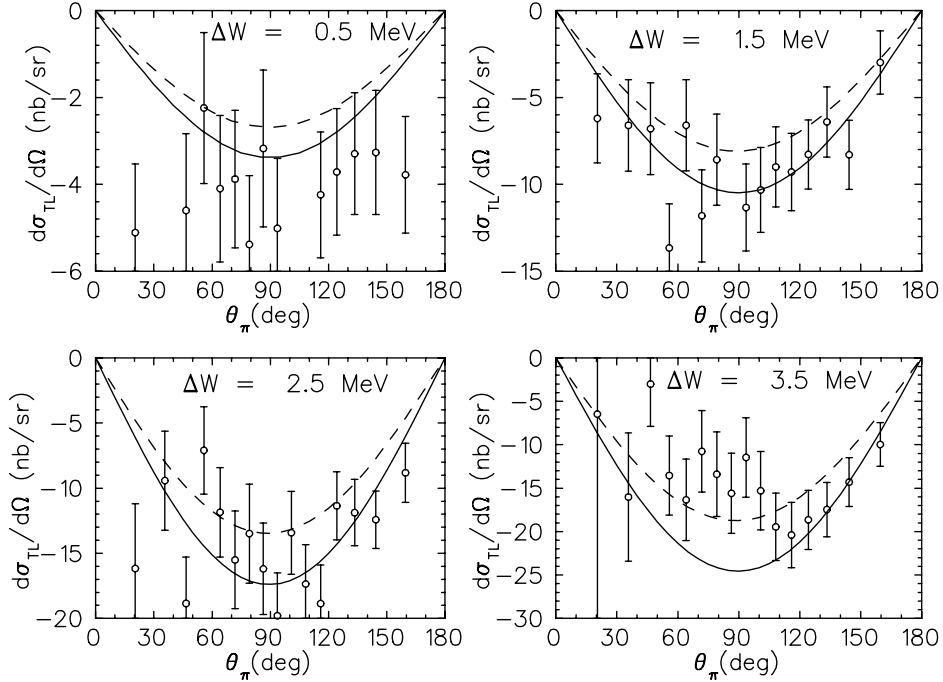, width=9 cm, angle=90} }
\caption{ Same as in Fig. 6 for the transverse-longitudinal cross
section $d\sigma_{TL}/d\Omega$.}
\end{figure}

\end{document}